# Spin filtering with insulating altermagnets

Kartik Samanta[1,*], Ding-Fu Shao[2,†], and Evgeny Y. Tsymbal[1,‡]

[1] *Department of Physics and Astronomy & Nebraska Center for Materials and Nanoscience, University of Nebraska, Lincoln, Nebraska 68588-0299, USA*

[2] *Key Laboratory of Materials Physics, Institute of Solid-State Physics, HFIPS, Chinese Academy of Sciences, Hefei 230031, China*

Altermagnetic (AM) materials have recently attracted significant interest due to the non-relativistic momentum-dependent spin splitting of their electronic band structure which may be useful for antiferromagnetic (AFM) spintronics. So far, however, most research studies have been focused on AM *metals* which can be utilized in spintronic devices, such as AFM tunnel junctions (AFMTJs). At the same time, AM *insulators* have remained largely unexplored in the realm of AFM spintronics. Here, we propose to employ AM insulators (AMIs) as efficient spin-filter materials. By analyzing the complex band structure of rutile-type altermagnets $M$F$_2$ ($M$ = Fe, Co, Ni), we demonstrate that the evanescent states in these AMIs exhibit spin- and momentum-dependent decay rates resulting in a substantial momentum-dependent spin polarization of the tunneling current. Using a model of spin-filter tunneling across a spin-dependent potential barrier, we estimate the TMR effect in spin-filter magnetic tunnel junctions (SF-MTJs) that include two magnetically decoupled $M$F$_2$ (001) barrier layers. We predict a sizable spin-filter TMR ratio of about 150-170% in SF-MTJs based on AMIs CoF$_2$ and NiF$_2$ if the Fermi energy is tuned to be close to the valence band maximum. Our results demonstrate that AMIs provide a viable alternative to conventional ferromagnetic or ferrimagnetic spin-filter materials, potentially advancing the development of next-generation AFM spintronic devices.

The field of spintronics exploits the spin degree of freedom in solid-state electronic devices for information processing and storage [1]. Spintronics relies on electric currents carrying spin polarization that is used to detect or control the magnetic order parameter. Obtaining a sizable spin polarization is therefore an important aspect of spintronics. One of the promising approaches to generate highly spin-polarized electric currents is to employ spin-filter tunneling, where a spin-filter material, typically a ferromagnetic (FM) insulator, is used as a barrier in a spin-filter magnetic tunnel junction (SF-MTJ) [2]. Spin-filter tunneling is considered as a viable alternative to a more conventional approach based on magnetic tunnel junctions (MTJs) which utilize the magnetic alignment of two FM electrodes to produce a tunneling magnetoresistance (TMR) effect [3-8].

Electron tunneling through an insulating layer can be understood in terms of evanescent states which represent the eigen-states of the insulator within its energy band gap [9]. Their wave-functions decay exponentially with a rate $\kappa$ that is determined by the complex band structure of the insulator [10-12]. In FM insulators, the exchange splitting of the spin bands leads to a spin-dependent tunneling barrier and hence spin-dependent decay rates. As tunneling transmission depends exponentially on the decay rate, electrons are transmitted with significantly different probabilities depending on their spin orientation, resulting in tunneling spin polarization [13]. This represents the basic idea of spin-filter tunneling.

A Tedrow-Meservey technique of spin-polarized tunneling into superconductors [14] has been first used to demonstrate spin-filter tunneling in FM Eu chalcogenides, such as EuS [15], EuSe [16], and EuO [17]. More recently, SF-MTJs have been utilized to explore a spin-filter TMR effect controlled by the relative magnetization orientation of an FM barrier layer and an FM counter electrode. Complex FM (or ferrimagnetic) oxide materials such as CoFe$_2$O$_4$ [18, 19], NiFe$_2$O$_4$ [20], NiMn$_2$O$_4$ [21,22], CoCr$_2$O$_4$ and MnCr$_2$O$_4$ [23] have been used in these experiments. Very recently 2D van der Waals materials, such as CrI$_3$ [24-26], CrBr$_3$ [27], and CrSe$_3$ [28], have been explored as FM barriers in SF-MTJs and demonstrated large spin-filter TMR effects.

In recent years, the emphasis of spintronics has been shifted from ferromagnets to antiferromagnets due to their advantages such as robustness against magnetic perturbations, absence of stray fields, and ultrafast spin dynamics [29-33]. Especially interesting appeared to be antiferromagnetic (AFM) crystals breaking $PT$ and $U\tau_{½}$ symmetries (where $P$ is space inversion, $T$ is time reversal, $U$ is spin-flip, and $\tau_{½}$ is half a unit cell translation) and thus producing momentum-dependent spin polarization, even in the absence of spin-orbit coupling (SOC) [34-43]. This subgroup of collinear antiferromagnets is now known as "altermagnets" [41,42]. Due to their spin-polarized band structure, altermagnets are predicted to exhibit spin-dependent transport properties [44]. Particularly interesting are AFM tunnel junctions (AFMTJs) with altermagnetic (AM) metals electrodes, who's relative Néel vector orientation controls the TMR effect [45-50].

While there has been an extensive endeavor to utilize AM metals in spintronic devices such as AFMTJs, AM insulators (AMIs) remained largely unexplored in the realm of AFM spintronics. It is notable, however, that among the predicted altermagnets, the majority (around 80%) are insulators [51]. It is therefore interesting to explore their spin-dependent properties in view of potential spintronic applications. One of the emerging questions, in this regard, is if AMIs could be used as efficient spin-filter barrier materials in SF-MTJs to generate highly spin-polarized currents and produce strong spin-filter TMR effects.

In this letter, we address this question by analyzing the complex band structure of rutile-type AMIs $M$F$_2$ ($M$ = Fe, Co, Ni). We demonstrate that evanescent states in these AMIs exhibit

spin- and momentum-dependent decay rates resulting in a large momentum-dependent spin polarization of the electric current tunneling across them. Based on these results, we estimate the TMR effect in SF-MTJs that consist of two $MF_2$ (001) barriers whose Néel vector can be switched to be parallel or antiparallel. We predict a sizable spin-filter TMR effect of about 150-170% in SF-MTJs based on AMI $CoF_2$ and $NiF_2$. Our results show that AMIs can be efficiently employed in spintronics as a viable alternative to conventional FM spin-filter materials.

Owing to the nonrelativistic spin splitting of AMIs, these materials exhibit evanescent states with different decay rates $\kappa^{\uparrow,\downarrow}(\vec{k}_\parallel)$ for up- (↑) and down- (↓) spin electrons. Their decay rates depend on the wave vector $\vec{k}_\parallel = (k_x, k_y)$ that is transverse to the transport direction and conserved in the tunneling process. Due to the exponential dependence of the transmission probability $\sim e^{-2\kappa^{\uparrow,\downarrow}(\vec{k}_\parallel)d}$, where $d$ is the barrier thickness, the transmitted electrons are spin polarized. We define the $\vec{k}_\parallel$-dependent spin-filtering factor of the AMI barrier as $f(\vec{k}_\parallel) = \kappa^\downarrow(\vec{k}_\parallel) - \kappa^\uparrow(\vec{k}_\parallel)$, which implies that up- (down-) spin electrons are transmitted more efficiently across the barrier for positive (negative) $f$. For a normal insulator, the spin-filtering factor $f(\vec{k}_\parallel) = 0$ [Fig. 1(b)], while for an AMI, $f(\vec{k}_\parallel) \neq 0$ [Fig. 1(c)]. If a non-spin polarized current from a normal metal (NM) [Fig. 1(a)] is transmitted across such AMI, it acquires $\vec{k}_\parallel$-dependent spin polarization $p(\vec{k}_\parallel) = \frac{e^{-2\kappa^\uparrow(\vec{k}_\parallel)d} - e^{-2\kappa^\downarrow(\vec{k}_\parallel)d}}{e^{-2\kappa^\uparrow(\vec{k}_\parallel)d} + e^{-2\kappa^\downarrow(\vec{k}_\parallel)d}} = \tanh[f(\vec{k}_\parallel)d]$, as illustrated in Fig. 1(d). In AMIs which host mirror symmetries, $M_x$ and $M_y$, such as $MnF_2$ (001) [52] and $CoF_2$ (001) [53], the spin-filtering factor obeys $M_x f(k_x, k_y) = -f(-k_x, k_y)$ and $M_y f(k_x, k_y) = -f(k_x, -k_y)$. As a result, $f(\vec{k}_\parallel)$ and $p(\vec{k}_\parallel)$ exhibit a $d$-wave shape with alternating sign in every quadrant of the (001) two-dimensional Brillouin zone (2DBZ) [Figs. 1(c,d)].

The $\vec{k}_\parallel$-dependent spin polarization of AFIs can be used to achieve spin-filter TMR in SF-MTJs representing two NM electrodes separated by a composite barrier layer [Figs. 1(e,f)]. The latter consists of two AMIs separated by a normal insulator to magnetically decouple the two AMIs. When the Néel vector is parallel [Fig. 1(e)], the spin-filtering factors $f(\vec{k}_\parallel)$ of the AMIs match at each $\vec{k}_\parallel$. This allows electrons with a given spin (either up or down) to be efficiently transmitted across both AMI layers in those regions of the 2DBZ where $|f(\vec{k}_\parallel)|$ is large, thus supporting a large tunneling current. On the contrary, when the Néel vector is antiparallel [Fig. 1(f)], the spin-filtering factors $f(\vec{k}_\parallel)$ of the AMIs mismatch. In this case, those electrons which have been efficiently transmitted across the first AMI layer are filtered out in the second AMI layer due to the spin mismatch, thus suppressing the tunneling current. As a result, the transmission probability $T_P(\vec{k}_\parallel)$ for the parallel-aligned SF-MTJ [Fig. 1(e)] is larger than the transmission probability $T_{AP}(\vec{k}_\parallel)$ for

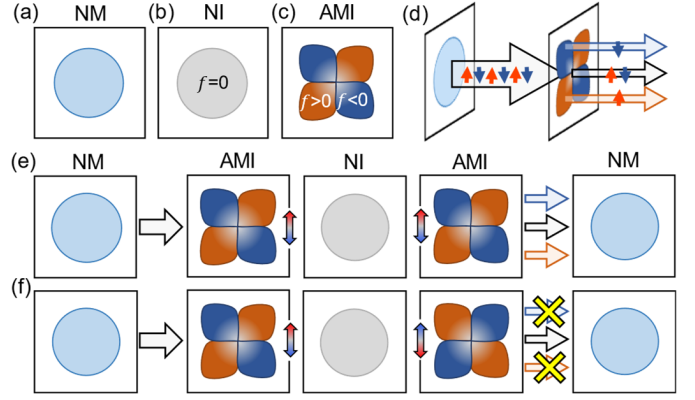

**Fig. 1** (a-c) Schematic views of normal metal (NM) with spin-independent Fermi surface (a), normal insulator (NI) with zero spin filtering factor (b), and altermagnetic insulator (AMI) with a non-zero $\vec{k}_\parallel$-dependent spin-filtering factor. (d) Schematic mechanism for the $\vec{k}_\parallel$-dependent tunneling spin polarization in a tunnel junction with a NM electrode and an AMI barrier. Block arrows indicate electric current. Red and blue arrows indicate up and down spins. (e,f) Schematic of SF-MTJ with a composite barrier which consists of two AMI layers separated by an NI layer for parallel (e) and antiparallel (f) Néel vector. Double arrows indicate the Néel vector in AMIs.

the antiparallel-aligned SF-MTJ [Fig. 1(f)]. This leads to a $\vec{k}_\parallel$-dependent spin-filter $TMR(\vec{k}_\parallel) = \frac{T_P(\vec{k}_\parallel) - T_{AP}(\vec{k}_\parallel)}{T_{AP}(\vec{k}_\parallel)}$ and a finite net spin-filter $TMR = \frac{T_P - T_{AP}}{T_{AP}}$, where $T_P = \frac{1}{N_k}\sum_{\vec{k}_\parallel} T_P(\vec{k}_\parallel)$, $T_{AP} = \frac{1}{N_k}\sum_{\vec{k}_\parallel} T_{AP}(\vec{k}_\parallel)$, and $N_k$ is the number of $k$-points.

To achieve an efficient spin-filtering effect, AMIs with a large spin splitting are required to support spin-dependent decay rates $\kappa^{\uparrow,\downarrow}$. Among AMIs, fluoride compounds are reported to have a sizable spin splitting [51]. Therefore, to demonstrate the spin-filtering effect, as representative AMIs, we consider rutile fluorides $MF_2$ ($M$ = Mn, Co, Ni) whose atomic structure is shown in Fig. 2(a) (space group P4$_2$/$mnm$ [54]). The magnetic moment of the metal atom $M1$ at the center of the unit cell is opposite to that of the $M2$ atoms at the corners. The Nèel vector is pointing along the [001] direction in $MnF_2$ [52] and $CoF_2$ [53], and along the [010] direction in $NiF_2$ [55].

We perform density-functional theory (DFT) calculations of the electronic and magnetic structure of bulk $MF_2$ crystals, as described in Supplementary Material [56]. These calculations predict that $MnF_2$, $CoF_2$, and $NiF_2$ have indirect band gaps of 4.14, 4.36 and 4.30 eV, respectively. The magnetic moments at the Mn, Co, Ni sites are found to be 4.95, 2.90, and 1.78 $\mu_B$, indicating a nominal 2+ valence of the $M$ ions with high-spin states exhibiting $d^5$ [$t_{2g}(3\uparrow)$, $e_g(2\uparrow)$], $d^6$ [$t_{2g}(3\uparrow, 2\downarrow)$, $e_g(2\uparrow)$], and $d^8$ [$t_{2g}(3\uparrow, 3\downarrow)$, $e_g(2\uparrow)$] occupancies, respectively. The calculated band gaps of the $MF_2$ compounds and their magnetic moments are in good agreement with the experimental observations [52,53,55] and the earlier calculations [36,51].

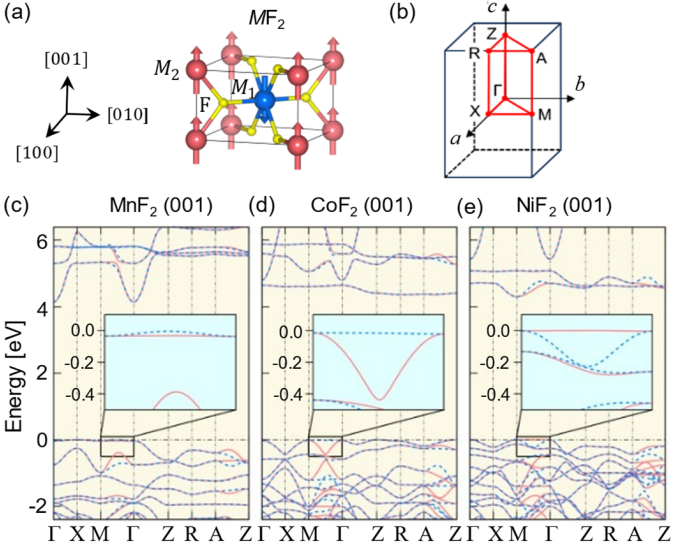

**Fig. 2** (a) Atomic and magnetic structure of AMI $MF_2$. (b) Tetragonal Brillouin zone of $MF_2$ with high-symmetry points and lines indicated. (c-e) Band structure of $MnF_2$ (c), $CoF_2$ (d), and $NiF_2$ (e) calculated without spin-orbit coupling. Red and blue curves indicate up- and down-spin bands, respectively. Zero energy is set to the VBM. Insets show the band structure in the vicinity of the VBM (-0.5 eV to 0.1 eV) along the Γ-M direction.

The magnetic space group of the $MF_2$ crystals has broken $PT$ or $U\tau_{1/2}$ symmetries and exhibits glide $G_x = \{M_x|(\frac{1}{2},\frac{1}{2},\frac{1}{2})\}$, $G_y = \{M_y|(\frac{1}{2},\frac{1}{2},\frac{1}{2})\}$ and mirror $M_z$ symmetries. In the absence of SOC, these symmetries enforce spin-degenerate bands at the $k$-planes invariant to $G_x$ and $G_y$, such as $k_x = 0, \pi/2$ and $k_y = 0, \pi/2$, whereas spin splitting is not constrained away from these planes. This is evident from Figs. 2(c-e), showing that the electronic bands of $MF_2$ are spin degenerate along the Γ-X, Γ-Z, Γ-Z, X-M, Z-R, and R-A directions lying in these glide-invariant planes, while they are spin-split along the Γ-M and Z-A directions, going away from these planes. We find a large spin splitting of about 0.44 eV in the vicinity of the VBM for $CoF_2$ [inset of Fig. 2(d)], whereas it is about 0.03 eV and 0.21 eV for $MnF_2$ and $NiF_2$, respectively [insets of Figs. 2(c,e)]. For all the $MF_2$ compounds, spin splitting around the conduction band minimum (CBM) is notably smaller than that near the VBM.

Due to this spin splitting, the $\vec{k}_\parallel$-dependent potential barrier for tunneling up- and down-spin electrons is different. This leads to spin-dependent evanescent states with different decay rates $\kappa^{\uparrow,\downarrow}$ for electrons with opposite spin. The evanescent states in insulators are determined by the complex band structure in the energy-gap region [10-13]. To obtain the complex band structure, we calculate the wave vector $k_z$ parallel to the transport direction as a function of $\vec{k}_\parallel$ and energy $E$ using the dispersion relation $E = E(\vec{k}_\parallel, k_z^{\uparrow,\downarrow})$ allowing complex solutions for $k_z^{\uparrow,\downarrow} = q^{\uparrow,\downarrow} + i\kappa^{\uparrow,\downarrow}$. The imaginary part $\kappa^{\uparrow,\downarrow}$ is the spin-dependent decay rate that determines the wave function $\sim e^{-\kappa^{\uparrow,\downarrow} d}$. The evanescent states with the lowest decay rate are expected to be most efficient in the tunneling process and hence control transmission.

Figs. 3(a-c) show the calculated complex band structure of $CoF_2$ (001) at different $\vec{k}_\parallel$-points [for $MnF_2$ and $NiF_2$, see Figs. S1(a-c) and S2(a-c)]. These complex bands are connected to the real bands with the same curvature at the connecting points due to the analytic properties of the $E(\vec{k}_\parallel, k_z^{\uparrow,\downarrow})$ function [Figs. S3-S5]. The complex bands at energies in the band gap of $MF_2$ (001) determine the evanescent states of interest. It seen that at $\vec{k}_\parallel = (0,0)$, i.e. at the $\bar{\Gamma}$ point, the complex bands are spin degenerate [Figs. 3(a), S1(a), and S2(a)], due to the spin degeneracy of the real bands along the Γ-Z line by symmetry [Figs. 2(c-e)]. Away from the $\bar{\Gamma}$ point (and $\bar{M}$ points) in the 2DBZ, the complex bands are spin-split, as seen from Figs. 3(b,c) for $CoF_2$ and Figs. S1(b,c) and S2(b,c) for $MnF_2$ and $NiF_2$, reflecting the spin splitting of the real bands away from the $G_{x,y}$ invariant planes [Figs. 2(c-e)]. For example, in $CoF_2$ (001), at $E = E_{VBM} + 0.1$ eV and $\vec{k}_\parallel = (0.3, 0.3)$, the lowest decay rates are $\kappa^\uparrow = 0.36$ Å$^{-1}$ and $\kappa^\downarrow = 0.23$ Å$^{-1}$ [indicated by blue and red circles in Fig. 3(b)].

Figs. 3(d-f) show the calculated distribution of the lowest decay rates $\kappa^{\uparrow,\downarrow}(\vec{k}_\parallel)$ in the 2DBZ for different energies within the band gap of $CoF_2$ (001) [for $MnF_2$ (001) and $NiF_2$ (001), see Figs. S1(d-f) and S2(d-f)]. We observe a sizable spin dependence of $\kappa^{\uparrow,\downarrow}(\vec{k}_\parallel)$, especially along the $\bar{\Gamma}$-$\bar{M}$ direction for energies close to the VBM, e.g. $E = E_{VBM} + 0.1$ eV [Fig. 3(d)]. This reflects a sizable $f(\vec{k}_\parallel)$ in $CoF_2$ [Fig. S6(d)], exceeding that in $MnF_2$ and $NiF_2$ [Figs. S6(a,g)]. For energies close to the middle of the gap (MG), $E = E_{MG}$, the decay rates become nearly spin independent [Figs. 3(e), S1(e) and S2(e)]. This reflects nearly vanishing $f(\vec{k}_\parallel)$ [Figs. S6(b,e,h)] due to the nearly spin degenerate complex bands at the midgap energy $E_{MG}$ of $MF_2$ (001) [Figs. 3(b), S1(b), and S2(b)]. For energies close to the CBM, spin dependence of $\kappa^{\uparrow,\downarrow}(\vec{k}_\parallel)$ and $f(\vec{k}_\parallel)$ is less pronounced than that for the energies close to the VBM for all three compounds $CoF_2$, $MnF_2$, and $NiF_2$ [Figs. 3(f), S1(f), S2(f), and S6(c-i)]. For example, at $E = E_{CBM} - 0.1$ eV, the lowest decay rates in $CoF_2$ (001) are $\kappa^\uparrow = 0.37$ Å$^{-1}$ and $\kappa^\downarrow = 0.41$ Å$^{-1}$ [indicated by blue and red circles in Fig. 3 (c)].

The spin-dependent decay rates $\kappa^{\uparrow,\downarrow}(\vec{k}_\parallel)$ and non-zero spin-filtering factors $f(\vec{k}_\parallel)$ in AMIs $MF_2$ lead to $\vec{k}_\parallel$-dependent spin polarization $p(\vec{k}_\parallel) = \tanh[f(\vec{k}_\parallel) d]$. For example, at $E = E_{VBM} + 0.1$ eV and barrier thickness $d = 2$ nm, $|p(\vec{k}_\parallel)|$ is sizable in a wide region of the 2DBZ [Figs. 4(a-c)]. Especially large $|p(\vec{k}_\parallel)|$ is in $CoF_2$ and $NiF_2$ approaching 100% at certain $\vec{k}_\parallel$ for energies close to the VBM. We also observe a sizable $|p(\vec{k}_\parallel)|$ at $E = E_{CBM} - 0.1$ eV, though in much narrower regions of the 2DBZ [Fig. S7].

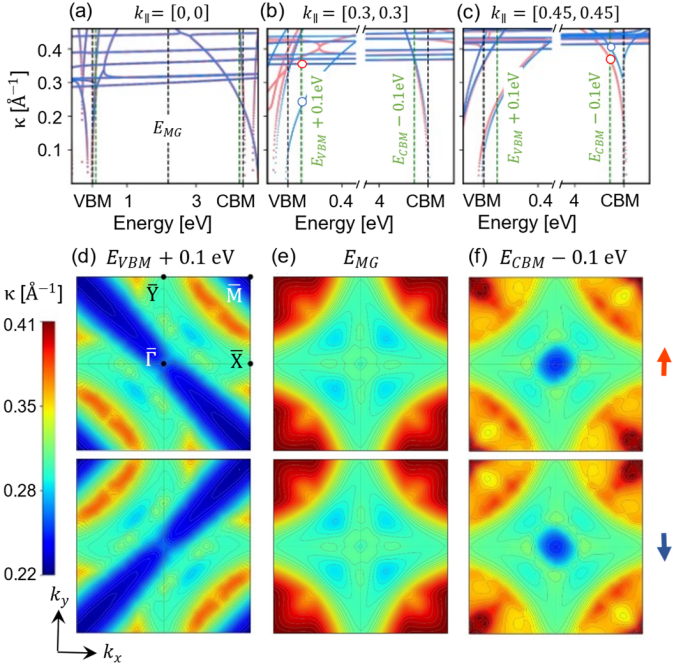

**Fig. 3** (a-c) Complex band structures of CoF$_2$ (001) as a function of energy at three different $\vec{k}_\parallel = (k_x, k_y)$ along the $\bar{\Gamma}$-$\bar{M}$ line. (d-f) Distribution of lowest decay rates for up-spin (upper panels) and down-spin (lower panels) evanescent states in the band gap of CoF$_2$ (001) as a function of $\vec{k}_\parallel = (k_x, k_y)$ in the 2DBZ at (d) $E = E_{VBM} + 0.1$ eV, (e) $E = E_{MG}$, and (f) $E = E_{CBM} - 0.1$ eV.

The predicted $\vec{k}_\parallel$-dependent spin polarization $p(\vec{k}_\parallel)$ can be functionalized in SF-MTJs to produce spin-filter TMR [Figs. 1(e,f)]. To estimate the TMR ratio, we employ a model of spin-filter tunneling through a double potential barrier whose spin-dependent energy profile depends on the relative Néel vector orientation of the two AMI layers. For simplicity, we assume $\vec{k}_\parallel$-independent conduction channels in the NM electrodes and ignore contribution from the non-magnetic spacer layer separating AMIs. Assuming equal thickness $d$ of the two AMI layers, we estimate tunneling transmission for parallel- [$T_P(\vec{k}_\parallel)$] and antiparallel- [$T_{AP}(\vec{k}_\parallel)$] aligned SF-MTJ, as follows:

$$T_P(\vec{k}_\parallel) \propto \left[ e^{-2\kappa^\uparrow(\vec{k}_\parallel)d} e^{-2\kappa^\uparrow(\vec{k}_\parallel)d} + e^{-2\kappa^\downarrow(\vec{k}_\parallel)d} e^{-2\kappa^\downarrow(\vec{k}_\parallel)d} \right], \quad (1)$$

$$T_{AP}(\vec{k}_\parallel) \propto \left[ e^{-2\kappa^\uparrow(\vec{k}_\parallel)d} e^{-2\kappa^\downarrow(\vec{k}_\parallel)d} + e^{-2\kappa^\downarrow(\vec{k}_\parallel)d} e^{-2\kappa^\uparrow(\vec{k}_\parallel)d} \right], \quad (2)$$

where $\kappa^{\uparrow,\downarrow}(\vec{k}_\parallel)$ are lowest decay rates obtained from the complex band structure. Figs. 4(d-f) and 4(g-i) show the calculated $T_P(\vec{k}_\parallel)$ and $T_{AP}(\vec{k}_\parallel)$ for $E = E_{VBM} + 0.1$ eV and $d = 2$ nm. It is seen that $T_P(\vec{k}_\parallel)$ exhibits a pattern resembling an overlap of the distribution patterns of $\kappa^\uparrow(\vec{k}_\parallel)$ and $\kappa^\downarrow(\vec{k}_\parallel)$ [e.g., compare Figs. 4(e) and 3(d)]. While $T_{AP}(\vec{k}_\parallel)$ also exhibits a qualitatively similar pattern, it features greatly suppressed transmission in the 2DBZ regions with the strongest contrast between $\kappa^\uparrow(\vec{k}_\parallel)$ and $\kappa^\downarrow(\vec{k}_\parallel)$ [e.g., around the middle of the $\bar{\Gamma}$-$\bar{M}$ lines in Figs. 4(g,i)]. Figs. 4(j-l) show the calculated $\vec{k}_\parallel$-resolved TMR ratio, $TMR(\vec{k}_\parallel) = \frac{T_P(\vec{k}_\parallel) - T_{AP}(\vec{k}_\parallel)}{T_{AP}(\vec{k}_\parallel)}$, at $E = E_{VBM} + 0.1$ eV. Giant contributions to TMR close to $10^4\%$ are seen from those regions in the 2DBZ where the spin polarization is approaching 100% [Figs. 4(a-c)]. This reflects the fact that $TMR(\vec{k}_\parallel)$ and $p(\vec{k}_\parallel)$ are related by Julliere's formula [3], $TMR(\vec{k}_\parallel) = \frac{2p^2(\vec{k}_\parallel)}{1-p^2(\vec{k}_\parallel)}$. Fig. 5 shows the total TMR ratio as a function of the Fermi energy. For all three compounds, we observe a strong enhancement of spin-filtering TMR at energies close to VBM with values of about 150–170 % for CoF$_2$ and NiF$_2$.

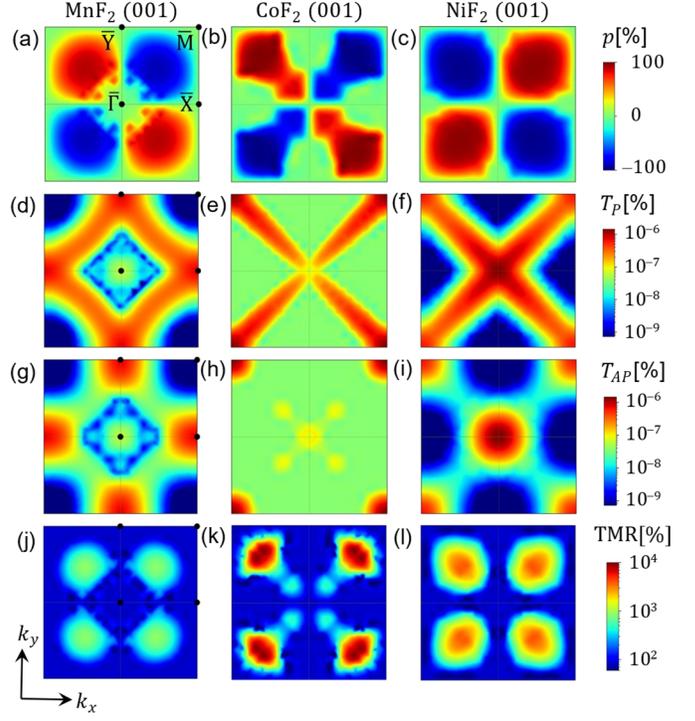

**Fig. 4** Transport properties of MnF$_2$ (a,d,g,j), CoF$_2$ (b,e,h,k) and NiF$_2$ (c,f,i,l) AMIs at $E = E_{VBM} + 0.1$ eV. (a-c) $\vec{k}_\parallel$-dependent spin polarization $p(\vec{k}_\parallel) = \tanh[f(\vec{k}_\parallel)d]$. (d-i) $\vec{k}_\parallel$-dependent transmission across an SF-MTJ for parallel (d-f) and antiparallel (g-i) Néel vectors. (j-l) $\vec{k}_\parallel$-resolved $TMR(\vec{k}_\parallel) = \frac{T_P(\vec{k}_\parallel) - T_{AP}(\vec{k}_\parallel)}{T_{AP}(\vec{k}_\parallel)}$. In (e) and (h), $T_P(\vec{k}_\parallel)$ and $T_{AP}(\vec{k}_\parallel)$ are scaled by a factor of 50 for a better view.

Our results demonstrate that AMIs $M$F$_2$ can be used as spin-filter materials in SF-MTJs provided that the Fermi energy can be tuned close to the VBM of the AMI. This can be achieved, e.g., by utilizing polar interfaces which allow varying the band offset between metals and insulators in a broad range of energies (see, e.g., Refs. [57,58]). To magnetically decouple two AMIs, one can use a non-magnetic insulator ZnF$_2$ which has the rutile structure and a very similar lattice constant [54].

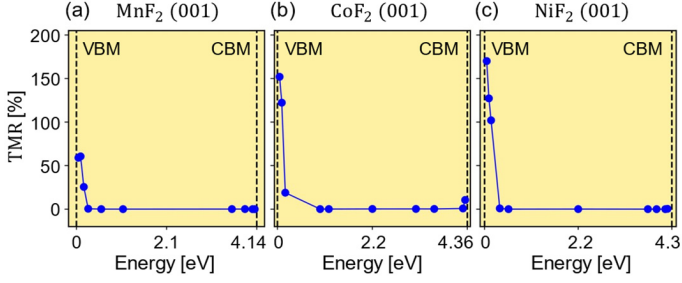

**Fig. 5** Spin-filter $TMR = \frac{T_P - T_{AP}}{T_{AP}}$ as a function of the Fermi energy for SF-MTJs based on MnF$_2$ (001) (a), CoF$_2$ (001) (b), and NiF$_2$ (001) (c).

We note that due to the mirror symmetries, $\hat{M}_x$ and $\hat{M}_y$, the net tunneling spin polarization $p = \sum_{\vec{k}_\parallel} p(\vec{k}_\parallel)$ of $M$F$_2$ (001) is zero for transport along the [001] direction. To obtain a finite net spin polarization, layer stacking with a low-symmetry surface is needed to break these symmetries. This can be achieved for transport along the [110] direction. Indeed, our calculations find a large net spin polarization for MnF$_2$ (110) at $E = E_{VBM} + 0.1$ eV [Figs. S8 and S9]. This property allows using AMIs as FM insulators in SF-MTJs.

Overall, our results indicate that AMIs can be efficiently employed in AFM spintronics to achieve strong spin-filtering effects. Using rutile fluorides $M$F$_2$ ($M$ = Mn, Co, Ni), as representative examples, we have predicted that these AMIs exhibit a large $\vec{k}_\parallel$-dependent spin polarization when the Fermi energy is tuned close to the VBM. This property allows realizing sizable spin-filter TMR in SF-MTJs based on these AMIs. Apart from rutile fluorides, there are numerous AMIs which may exhibit similar spin-filtering properties. We therefore hope that our results will motivate experimentalists working in this field to verify our predictions.

**Acknowledgments.** This work was supported by the National Science Foundation (Grant No. DMR-2316665) and UNL's Grand Challenges catalyst award "Quantum Approaches Addressing Global Threats." D.-F.S. acknowledges support from the National Key R&D Program of China (Grant No. 2021YFA1600200), the National Natural Science Foundation of China (Grants Nos. 12274411, 12241405, and 52250418), the Basic Research Program of the Chinese Academy of Sciences Based on Major Scientific Infrastructures (Grant No. JZHKYPT-2021-08), and the CAS Project for Young Scientists in Basic Research (Grant No. YSBR-084). Computations were performed at the University of Nebraska Holland Computing Center.

* ksamanta2@unl.edu
† dfshao@issp.ac.cn
‡ tsymbal@unl.edu